\documentclass[aps,superscriptaddress,floats,showpacs]{revtex4}
\usepackage{bm}
\usepackage{graphicx}
\usepackage{subfigure}

\begin{document}
\title{Buoyancy statistics in moist turbulent Rayleigh-B\'{e}nard convection}
\author{J\"org Schumacher}
\affiliation{Institut f\"ur Thermo- und Fluiddynamik, Technische Universit\"at Ilmenau, 
Postfach 100565, D-98684 Ilmenau, Germany}
\author{Olivier Pauluis}
\affiliation{Courant Institute of Mathematical Sciences, New York University, 
251 Mercer Street, New York, NY 10012-1185, USA}
\date{\today}

\begin{abstract}
We study shallow moist Rayleigh-B\'{e}nard convection in the Boussinesq approximation
in three-dimensional direct numerical simulations. The thermodynamics of phase changes 
is approximated by a piecewise linear equation of state close to the phase boundary. The impact of phase
changes on the turbulent fluctuations and the transfer of buoyancy through the layer is discussed as a
function of the Rayleigh number and the ability to form liquid water. The enhanced buoyancy flux due
to phase changes is compared with dry convection reference cases and related to the cloud cover in the 
convection layer. This study indicates that the moist Rayleigh-B\'{e}nard problem offers a practical framework
for the development and evaluation of parameterizations for atmospheric convection.
\end{abstract}

\maketitle

\section{Introduction}   
Moist thermal convection combines turbulent convection with phase changes and
latent heat release. It is ubiquitous throughout the atmosphere of the Earth 
(Heintzenberg \& Charlson 2009).   When a parcel of air rises in convective 
motion, it expands adiabatically. As a consequence, its temperature and 
pressure drop and at some point during its ascent the air parcel becomes 
saturated. Once water condenses, a cloud is formed. The range of spatial  and 
temporal scales in the convective turbulent motion varies widely, from a few 
hundred meters in isolated cumulus clouds to several thousands of kilometers 
in midlatitudes storm systems.   

Despite its enormous importance, the small-scale structure and statistics of moist  
convective turbulence has  been studied relatively little compared to its dry convection
counterpart.  The reason for this gap is that turbulent convection in moist air includes 
the complex nonlinear thermodynamics of phase changes in addition to the 
turbulent motion (Stevens 2005; Pauluis 2008). The associated latent heat release  
provides a rapidly changing 
local source of buoyant motion, so that moist convection is characterized by 
a complex interaction between dynamics and  thermodynamics.
One approach to this problem is to express the buoyancy of a parcel of moist air as function of 
its entropy, pressure and total water content. In such framework, 
phase changes can be treated implicitly, and lead to discontinuities 
of the partial derivatives in the equation of state at the saturation point (Emanuel 1994).  
While moist convection remains poorly understood, 
significant progress has been made in the last decade in understanding 
the global and local mechanisms of turbulent heat transfer in dry convection 
(for a comprehensive review see Ahlers {\it et al.} 2009).   
In this work, we aim at transferring some of the numerical analysis concepts from the well-investigated 
dry convection case, such as studies of the  Rayleigh number dependence of the heat transfer 
(Verzicco \& Camussi 2003), the flow properties in the cell (van Reeuwijk {\it et al.} 2008)
or the small-scale statistics (Emran \& Schumacher 2008) to the less-explored moist  convection 
case. 

We propose here to take a first step by considering moist convection in the idealized setting 
of moist Rayleigh-B\'{e}nard convection with a linearized  thermodynamics of phase changes 
(Pauluis \& Schumacher 2010). On the one hand, the model is a straightforward extension
of numerical studies in  dry Rayleigh-B\'{e}nard convection in the Boussinesq approximation (e.g. Schumacher 
2009). On the other hand, it is  a generalization of a moist convection model which was discussed 
by Bretherton (1987, 1988) for the linear and weakly nonlinear regime and has not been studied ever 
since. Here, we conduct direct 
numerical simulations of the turbulent nonlinear stage of moist convection. The present work reports 
systematic parameter investigations to understand the effect of phase change on the turbulent transport 
of buoyancy through the shallow layer. We also discuss the dependence of the cloud cover on the
physical parameters of the model.

\section{Moist Boussinesq model}
The buoyancy $B$ in atmospheric convection is given by (Emanuel 1994) 
\begin{equation}
B(S,q_v,q_l,q_i,p)=-g \frac{\rho(S,q_v,q_l,q_i,p)-\overline{\rho}}{\overline{\rho}} \,,
\label{buo0}
\end{equation}
with $g$ being the gravity acceleration, $\overline{\rho}$ a mean density, $p$ the pressure, 
$S$ the entropy and $q_v$, $q_l$, $q_i$ the mixing ratios of water vapor, liquid water and ice. 
In the following, we discuss in brief the sequence of simplifications of the equation of state 
that result in a model of shallow non-precipitating moist convection in the Boussinesq 
approximation -- the simplest case that goes beyond the well-known dry convection (Pauluis \& Schumacher 2010). First, in the Boussinesq approximation the pressure variations about a mean 
hydrostatic profile are omitted when computing the buoyancy (Pauluis 2008) and  one 
is left with $B(S,q_v,q_l,q_i,z)$.  Second, warm clouds are discussed with $q_i=0$. 
Third, we assume that the air parcels are in  local thermodynamic equilibrium, which means 
that water vapor and condensed water can only co-exist at saturation line. This implies that liquid 
water is formed whenever a relative humidity of 100\% is exceeded. Furthermore, no 
rain can fall out in our model.  The two remaining mixing ratios are then combined to 
the total water mixing ratio,  $q_T=q_v+q_l$. This assumption also excludes the possibility of supersaturation. Condensation in the Earth's atmosphere occurs primarily through heterogenous nucleation caused by a large number of cloud condensation nuclei ($n \sim 10^9$ m$^{-3}$). As a consequence, supersaturation rarely exceeds one per cent (Rogers \& Yau, 1989). 
In the absence of condensation nuclei in the fluid, homogeneous condensation may result in much larger supersaturation, as in a recent experiment by Zhong et al. (2009) where the condensate
is formed at the top plate of the convection cell. The assumption of local thermodynamic equilibrium 
has the practical advantage that, once the entropy and pressure are known, the total water content 
can be separated between the vapor and liquid phases. The dependencies of the buoyancy are thus reduced to  $B(S,q_T,z)$.  The buoyancy  is still a highly nonlinear function of the entropy, total 
water mixing ratio and height. Fourth, we approximate $B$ as a piecewise linear function of the two 
state variables $S$, $q_T$ at each height $z$ around the phase boundary between gas and liquid. 
The linearization step restricts us to a shallow layer since the height variations of thermodynamic quantities have to remain small.  It preserves the main physical ingredient: the discontinuity of partial derivatives (e.g. the specific heat) at the phase boundary. It also allows for an explicit determination 
of whether an air parcel is saturated or not.  Finally, since $B$ is a linear function of $S$ and $q_T$, 
we can introduce two new prognostic buoyancy fields, a {\em dry buoyancy field} $D$ (which corresponds to a liquid water potential temperature) and a {\em moist buoyancy field} 
$M$ (which corresponds to an equivalent potential temperature).  They are linear combinations 
of $S$ and $q_T$.  Since the state variables $S$ and $q_T$ are adiabatic invariants, the two new 
state variables $M$ and $D$ are also conserved by adiabatic transformations.
Consequently, the original buoyancy $B(S,q_v,q_l,q_i,p)$ is simplified to $B(M,D,z)$, 
a linear function of the fields $M$ and $D$ which is given by       
\begin{equation}
B({\bf x},t)=\max\left( M({\bf x},t), D({\bf x},t)-N^2_s z \right) \,,
\label{buo1}
\end{equation}
where $N_s$ is the Brunt-Vaisala frequency. This is the saturation condition in our model.

The dry and moist buoyancy fields can be decomposed in
\begin{eqnarray}
D({\bf x},t) &=& \overline{D}(z)+D^{\prime}({\bf x},t)=D_0+\frac{D_H-D_0}{H}z+D^{\prime}({\bf x},t)
\label{decomposition a}\\
M({\bf x},t) &=& \overline{M}(z)+M^{\prime}({\bf x},t)=M_0+\frac{M_H-M_0}{H}z+M^{\prime}({\bf x},t)\,.
\label{decomposition b}
\end{eqnarray}
The variations about the mean linear profiles of both fields have to vanish at $z=0$ and $H$. 
Equation (\ref{buo1}) can now be transformed into
\begin{equation}
B=\overline{M}(z) + \max\left( M', D' + \overline{D}(z) -\overline{M}(z)  -N^2_s z \right) \,.
\label{buo2}
\end{equation}
Note that the first term on the right-hand side is horizontally uniform. 
This implies that it can be balanced by a horizontally uniform pressure 
field given by $p(z) = -M_0 z -[(M_H - M_0)/(2H)] z^2$. We can thus 
remove the mean contribution from the  buoyancy field  without any loss of generality. 
A dimensionless 
version of the equations of motion is obtained by defining the characteristic quantities.
These are the height of the layer $H$, the free-fall velocity $U_f=\sqrt{H (M_0-M_H)}$, the time
$T_f=H/U_f$, the characteristic pressure $U_f^2$, and the buoyancy difference $M_0-M_H$.
The equations together with the decompositions (\ref{decomposition a}) and
(\ref{decomposition b}) are given by  
\begin{eqnarray}
\frac{\partial{\bf u}}{\partial t}+(\bf{u\cdot\nabla})\bf u &=& - \nabla p + \sqrt{\frac{Pr}{Ra_M}} \nabla^2{ \bf u} + B(M, D, z) {\bf e}_z \label{ueq} \\
\nabla \cdot \bf {u} &= &0 \label{ceq} \\
\frac{\partial D^{\prime}}{\partial t}+ ({\bf u}\cdot\nabla) D^{\prime}& =&  \frac{1}{\sqrt{Pr Ra_M}}  \nabla^2 D^{\prime} +\frac{Ra_D}{Ra_M} u_z 
\label{deq}\\
\frac{\partial M^{\prime}}{\partial t}+ ({\bf u}\cdot\nabla) M^{\prime}& =&  \frac{1}{\sqrt{Pr Ra_M}}  \nabla^2 M^{\prime} +u_z 
\label{meq}
\end{eqnarray}
These equations contain three non-dimensional parameters, the Prandtl number $Pr$, the dry
and the moist Rayleigh numbers $Ra_D$ and $Ra_M$  
\begin{equation}
Pr =  \frac{\nu}{\kappa}, \;\;\;\;\;Ra_D =\frac{H^3 (D_0 - D_H )}{\nu \kappa},  
\;\;\;\;\;Ra_M =\frac{H^3 (M_0 - M_H )}{\nu \kappa}\,.
\end{equation}
Under most circumstances, the amount of water in the atmosphere decreases with height. 
This implies that the moist Rayleigh number should be larger than the dry Rayleigh number, 
$Ra_M \ge Ra_D$. In addition to the three parameters explicitly present in equations 
(\ref{ueq})-(\ref{meq}), two more parameters are hidden implicitly within the definition (\ref{buo2})
of the buoyancy $B$ which is given in dimensionless form by 
\begin{equation}
B = \max \left( M^{\prime}, D^{\prime} +  SSD +  
\left(1-\frac{Ra_D}{Ra_M}\right) z  - CSA z \right) \,.
\label{bstar}
\end{equation}
The so-called {\it Surface Saturation Deficit} 
$SSD$ and the {\it Condensation in Saturated Ascent} $CSA$ are then defined as
\begin{equation}
SSD = \frac{D_0 - M_0}{M_0 - M_H} \;\;\; \text{and}\;\;\; 
CSA = \frac{N^2_s H}{M_0 - M_H}\,. \label{CSA} 
\end{equation}
These two new non-dimensional parameters respectively measure how close 
the lower  boundary is to saturation, and how much water can condense within 
the atmospheric layer during an adiabatic ascent of a saturated air parcel. The larger $CSA$, the 
easier is the formation of liquid water and thus of clouds. 
When $D_0 - M_0$ is positive, the air at the lower boundary 
is unsaturated, and $D_0 - M_0$ is proportional to the "water deficit", i.e. the amount of 
water vapor that must be added to the air parcel to become saturated.  A positive  
Surface Saturation Deficit $SSD$ would occur over the continents.  For convection 
over the ocean,  the lower boundary is neither saturated nor unsaturated, i.e. $SSD = 0$.  
It is clear that we can consider a subspace of the five-dimensional parameter space only 
which is spanned in general by $Ra_D, Ra_M, Pr, SSD$ and $CSA$. Therefore, this study 
is restricted to $Pr=0.7$ and  $SSD=0$. The variation of $SSD$ while 
keeping the other parameters fixed was discussed already in Pauluis \& Schumacher (2010).  

The equations of motion are solved by a pseudospectral scheme with volumetric 
fast Fourier transformations and 2/3 de-aliasing in a Cartesian slab with side lengths 
$\Gamma H\times \Gamma H\times H$.  Here $\Gamma$ is the aspect ratio of the slab. In lateral directions
$x$ and $y$, we apply periodic boundary conditions. In the vertical $z$ direction, we apply free-slip
boundary conditions,
\begin{equation}
u_z=D^{\prime}=M^{\prime}=0 \;\;\; \text{and}\;\;\; \frac{\partial u_x}{\partial z}= 
\frac{\partial u_y}{\partial z}=0\,.
\label{bc}
\end{equation}
The boundary conditions, which have also been used in Bretherton (1987, 1988),  approximate  a situation over
an ocean surface at the bottom and a temperature inversion at the top.  Time-stepping is done by a second-order
Runge-Kutta scheme. Since both buoyancy fields are linearly unstable, the requirements
on mesh resolution and time stepping are the same as in dry convection. The additional scalar field and the 
update of the $B$ increases computational costs by 20\%. Table 1 summarizes the grid resolutions and dimensionless 
parameter sets which are taken in the direct numerical simulations. The spectral resolution does
not go below $k_{max}\eta_K=2.45$ for all DNS, where $k_{max}$ is the maximum resolved 
wavenumber and $\eta_K$ the Kolmogorov length.  Technically, we use $B^{\prime}$ in the momentum 
equation (\ref{ueq}) instead of $B$ since the mean contribution is $\overline{B}(z)$ which can be added
to the kinematic pressure, i.e. $\partial_z p+B=\partial_z \tilde{p}+B^{\prime}$.  
For the moist runs we 
distinguish two classes for initial equilibrium configurations -- a fully saturated slab which
corresponds with $\overline{M}(z)>\overline{D}(z)-N_s^2 z$ (large CSA) and a fully unsaturated  
slab with $\overline{M}(z)<\overline{D}(z)-N_s^2 z$ (small CSA).
\begin{table}
\begin{center}
\begin{tabular}{lccccccccc}
$Run$ & $N_x\times N_y\times N_z$ & $Ra_M$ & $Ra_D$ & $CSA$ & $U_f$ & $T_f$ & $T/T_f$ 
& $\frac{u_{rms}}{U_f}$ & $\frac{M_{rms}}{M_0-M_H}$\\
& & & & & \\ 
1$^{\ast}$ &$512\times 512\times 65$  & $9.5\times 10^5$ & $7.0\times 10^5$ & 0.53 & 3.06 & 1.03 & 141 & 0.356 & 0.434\\
2$^{\ast}$ &$512\times 512\times 65$  & $1.1\times 10^6$ & $7.0\times 10^5$ & 0.44 & 3.35 & 0.94 & 154 & 0.334 & 0.436\\
3 &$512\times 512\times 65$  & $1.4\times 10^6$ & $7.0\times 10^5$ & 0.35 & 3.75 & 0.84 & 173 & 0.272 & 0.432\\
4 &$512\times 512\times 65$  & $1.9\times 10^6$ & $7.0\times 10^5$ & 0.26 & 4.36 & 0.72 & 368 & 0.225 & 0.431\\
5 &$512\times 512\times 65$  & $2.9\times 10^6$ & $7.0\times 10^5$ & 0.17 & 5.35 & 0.59 & 449 & 0.184 & 0.433\\
6 &$512\times 512\times 65$  & --                             & $7.0\times 10^5$ & 0.00 & 2.63 & 1.19 & 122 & 0.362 & --\\
& & & & & \\
7$^{\ast}$ &$1024\times 1024\times 129$  & $9.5\times 10^6$ & $7.0\times 10^6$ & 0.53 & 3.06 & 1.03 & 342 & 0.308 & 0.436\\
8$^{\ast}$ &$1024\times 1024\times 129$  & $1.1\times 10^7$ & $7.0\times 10^6$ & 0.44 & 3.35 & 0.94 & 407 & 0.287 & 0.437\\
9 &$1024\times 1024\times 129$  & $1.4\times 10^7$ & $7.0\times 10^6$ & 0.35 & 3.75 & 0.84 & 412 & 0.240 & 0.436\\
10&$1024\times 1024\times 129$ &  $1.9\times 10^7$ & $7.0\times 10^6$ & 0.26 & 4.36 & 0.72 & 431& 0.194 & 0.434\\
11&$1024\times 1024\times 129$  & $2.9\times 10^7$ & $7.0\times 10^6$ & 0.17 & 5.35 & 0.59 & 587& 0.158 & 0.436\\
12&$1024\times 1024\times 129$  & --                             & $7.0\times 10^6$ & 0.00 & 2.63  & 1.19 & 321& 0.320 & --\\
& & & & & \\
13$^{\ast}$ &$2048\times 2048\times 257$  & $1.1\times 10^8$ & $7.0\times 10^7$ & 0.44 & 3.35 & 0.94 & 125 & 0.258 & 0.439\\
14 &$2048\times 2048\times 257$  & $1.9\times 10^8$ & $7.0\times 10^7$ & 0.26 & 4.36 & 0.72 & 150 & 0.176 & 0.438\\
\end{tabular}
\caption{Parameters of simulation runs:  grid resolution,  $Ra_M$, $Ra_D$ and $CSA$. For all runs, $Pr=0.7$, 
$\Gamma=8$ and $SSD=0$. We also display the characteristic velocity $U_f=\sqrt{(M_0-M_H) H}$, the characteristic 
time scale $T_f=H/U_f$ and the total integration time $T/T_f$. For dry runs 6 and 12, $U_f=\sqrt{(D_0-D_H) H}$. 
Furthermore, $u_{rms}/U_f$ with $u_{rms}=\sqrt{\langle u_x^2+u_y^2+u_z^2\rangle_{x,y,z,t}}$ and $M_{rms}/(M_0-M_H)$
with $M_{rms}=\sqrt{\langle M^{\prime\,2}\rangle_{x,y,z,t}}$ are shown. Runs that are labeled with an asterisk start out of a 
completely saturated equilibrium, $\overline{M}(z)>\overline{D}(z)-N_s^2 z$.} 
\end{center}
\label{tab1}
\end{table}

\section{Results}

\subsection{Buoyancy and velocity fluctuations}
\begin{figure}
\begin{center}
\includegraphics[width=8cm]{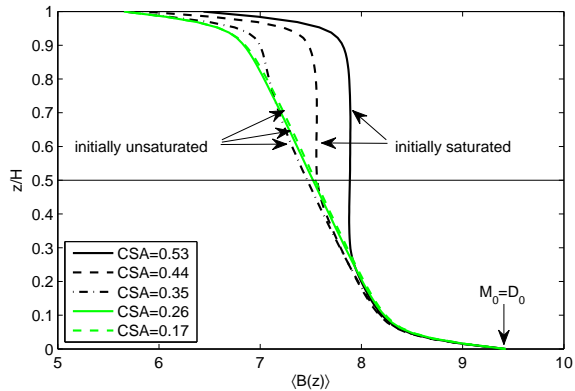}
\caption{Mean total buoyancy profiles $\langle B(z)\rangle$ for Runs 1 to 5. All runs have the 
same amplitude of $M_0$ (which equals $D_0$). It is also indicated for which runs the initial equilibrium
solution is completely unsaturated or saturated. Profiles for the series  with $Ra_D=7\times 10^6$
look qualitatively similar, except that the boundary layer thickness decreased.} 
\label{fig1}
\end{center}
\end{figure}
\begin{figure}
\begin{center}
\includegraphics[width=12cm]{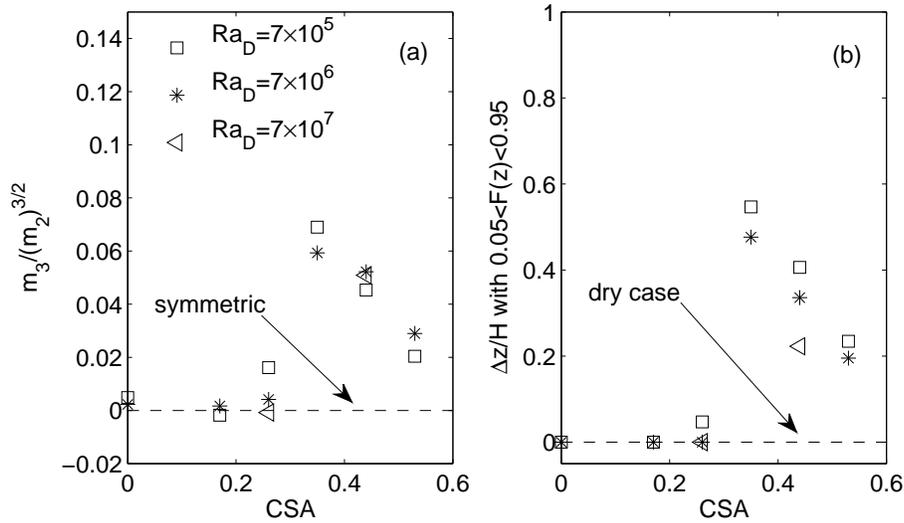}
\caption{Asymmetry of the velocity fluctuations as a function of $CSA$ and $Ra_D$. (a) Skewness 
$m_3/m_2^{3/2}$of the vertical profiles of the root mean square of the vertical velocity component $u_z$. (b) 
Vertical fraction $\Delta z/H$ of the layer with $0.05<F(z)<0.95$ where $F(z)$ is given by (\ref{fz}).} 
\label{fig2}
\end{center}
\end{figure}
\begin{figure}
\begin{center}
\includegraphics[width=14cm]{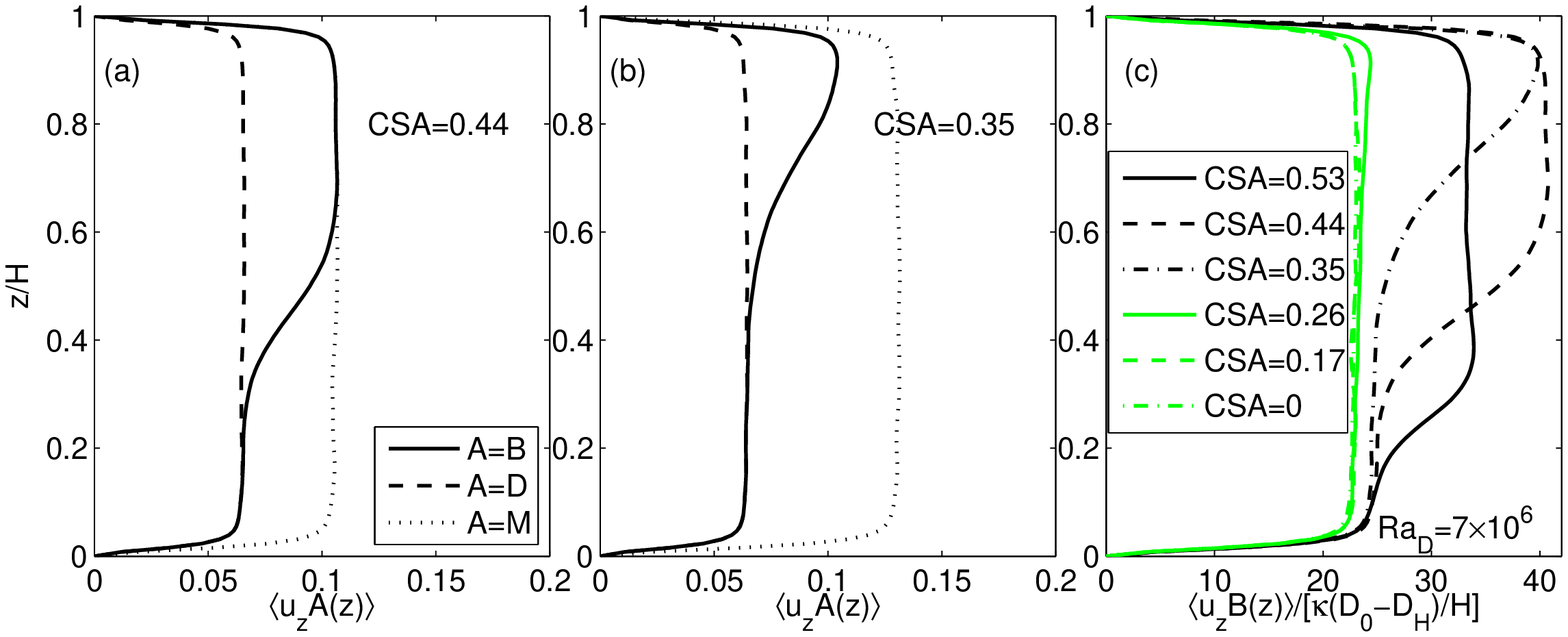}
\caption{Convective part of the buoyancy flux as a function of $Ra_D$, $Ra_M$ and $CSA$. 
(a) Relation between the fluxes  $\langle u_z D(z)\rangle$, $\langle u_z M(z)\rangle$ and $\langle u_z B(z)\rangle$. 
Data are for  $CSA=0.44$ and $Ra_D=7.0\times 10^6$.  (b)  Same as (a) for $CSA=0.35$. (c) 
$\langle u_z B(z)\rangle$ normalized by the corresponding dry diffusive buoyancy flux which is given as 
$\kappa (D_0-D_H)/H$. Data are for runs 7 to 12.} 
\label{fig3}
\end{center}
\end{figure}
Initially the equilibrium configuration is perturbed
infinitesimally and after $T/T_f\sim 10^2$ the flow is relaxed into a fully developed and 
statistically stationary turbulent state. This is when the statistical analysis is started. As indicated
in the table, we restrict dependencies of moist convection on the two Rayleigh numbers and parameter $CSA$.
Figure \ref{fig1} shows the mean vertical profile of 
the buoyancy, $\langle B(z)\rangle$, which is obtained by taking averages in the $z$ planes
and over an ensemble of  statistically independent snapshots.  It is observed, that this profile 
becomes strongly asymmetric for the cases which started with an initially fully saturated 
equilibrium. These will be the cases where phase changes affect the turbulence properties
most strongly.

Table 1 lists the root-mean-square (rms) values of $M$ and ${\bf u}$ as obtained in the statistically 
stationary regime. Since both buoyancy fields follow a linear advection-diffusion equation, the ratio of the 
rms fluctuations to the outer buoyancy difference $M_0-M_H$ should be constant. The velocity rms fluctuations 
decrease with decreasing $CSA$. At fixed $CSA$, the fluctuations  decrease also with increasing 
Rayleigh number $Ra_D$. This result is also observed in dry convection (Verzicco \& Camussi 2003). 
In Fig. \ref{fig2} we refine this analysis and study the vertical profile of the rms  
of the vertical velocity component, $u_{z,rms}(z)=\sqrt{\langle u_z^2(z)\rangle_{x,y,t}}$. A measure of 
asymmetry of the profile with respect to the midplane $z=H/2$ can be based on the moments 
$m_n=\int_0^H \left(z-\frac{H}{2}\right)^n u_{z,rms}(z)\,\mbox{d}z$. 
If the skewness $m_3/m_2^{3/2}$ is larger than zero then the vertical velocity fluctuations are 
enhanced in the upper half of the slab. Figure \ref{fig2} (a) shows that the profile is symmetric for the dry
reference run and those with smaller amount of water which can be condensed. Asymmetry is observed for 
$CSA\ge 0.35$ which peaks at $CSA=0.35$ and decreases again for larger $CSA$. We will show at the end of 
subsection 3.2 that the asymmetry in the vertical velocity fluctuations is directly coupled to the vertical fraction 
$\Delta z/H$ of the convection layer that is partially saturated and unsaturated. This fraction turns out to be
largest at $CSA=0.35$ for all $Ra_D$ (see Fig. \ref{fig2} (b)). It is also found 
that isotropy in the velocity fluctuations is 
established to a better degree with increasing $Ra_D$. We conclude
that phase changes cause the asymmetry of the vertical velocity fluctuations. However, with increasing 
Rayleigh number and thus Reynolds number the small-scale turbulence is found at increasingly isotropic 
conditions which can compensate this trend in parts.      

Of central importance in dry convection is the one-point-correlation between buoyancy (or temperature)
and vertical velocity, $\langle u_z B(z)\rangle$ (which is equal to $\langle u_z B^{\prime}(z)\rangle$). 
It enters the definition of the dimensionless measure of  buoyancy flux through the layer, 
the Nusselt number $Nu$. In the present model, we can define two Nusselt numbers for both 
fields in a standard way, such as  $Nu_D(z)=[ \langle u_z D(z)\rangle-\kappa \partial_z\langle D(z)\rangle]/
[\kappa(D_0-D_H)/H]$ for $D$ which is constant and thus simply denoted by $Nu_D$. Since $Nu_D$ 
and $Nu_M$ are normalized with respect to their diffusive fluxes, $Nu_D=Nu_M$ follows
which was verified in the simulations.  In order to quantify the additional amount of buoyancy 
transfer, we will relate the correlations $\langle u_z B(z)\rangle$ to the dry field in the following. 
Note that the buoyancy flux $\langle u_z B(z)\rangle $ is tied to the correlations $\langle u_zD(z)
\rangle$ and $\langle u_z M(z)\rangle$. Moreover, because the partial derivative of the buoyancy with 
respect to $M$ and $D$ are bound by 0 and 1 (see (\ref{buo1})), we have automatically that at each level $z$ 
\begin{equation}
\langle u_z D(z)\rangle\le \langle u_z B(z)\rangle \le \langle u_z M(z)\rangle\,.
\label{inequality}
\end{equation}
The lower bound ($\langle u_z D(z)\rangle= \langle u_z B(z)\rangle$) occurs when a layer is fully unsaturated, 
while the upper bound  ($\langle u_z B(z)\rangle= \langle u_z M(z)\rangle$) is achieved in fully saturated layer. 
This is demonstrated in Figs. \ref{fig3} (a) and (b). In Fig. \ref{fig3} (c),  we 
normalize the correlation by the dry diffusive buoyancy flux (which would correspond with $Nu_D=1$).
It is given by $\kappa (D_0-D_H)/H$. 
Again, we observe an enhancement of the correlations for the three largest values of $CSA$. 
The profiles for the two smaller values of $CSA$ collapse almost perfectly with the corresponding
dry reference cases in both series of DNS. Note that a normalization by the moist diffusive buoyancy flux
$\kappa (M_0-M_H)/H$ would result in systematic growth of the correlation since an 
increase of $Ra_M$ is in line with a decrease of $CSA$. Finally, the correlations 
increase as well when the Rayleigh numbers $Ra_D$ and $Ra_M$ are enhanced at given $CSA$. 
\begin{figure}
\begin{center}
\includegraphics[width=9cm]{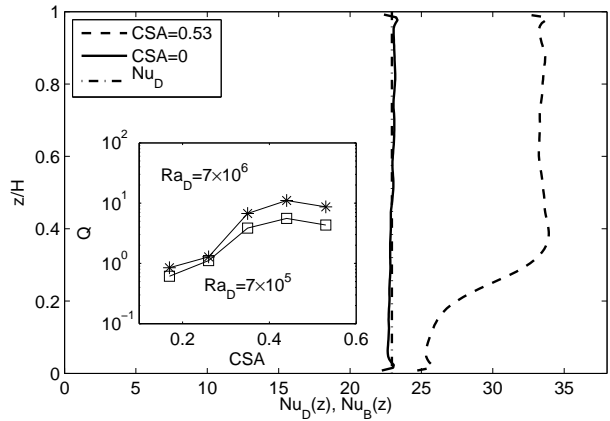}
\caption{Vertical profiles of $Nu_D(z)$ and $Nu_B(z)$ as given by Eq. (\ref{nud1}).
Data are for Runs 7 and 12. The additional amount of buoyancy, $Q$ is the area 
between moist and dry reference profiles. The inset shows $Q$ as a function of $CSA$ 
and $Ra_D$ (Runs 1--5 and 7--11).} 
\label{fig4}
\end{center}
\end{figure}

On the basis of the correlations between buoyancy and vertical velocity and the mean vertical profiles,
the additional buoyancy flux due to phase changes can be determined. 
We define  a Nusselt number based on the dry diffusive buoyancy flux:
\begin{equation}
Nu_B(z)=\frac{\langle u_z B(z)\rangle-\kappa\partial_z\langle B(z)\rangle}{\kappa (D_0-D_H)/H}\,,
\label{nud1}
\end{equation}
which is not necessarily constant with height, as can be seen in Fig. \ref{fig4}. Similar behaviour was found
by Oresta {\it et al.} (2009) in bubbly convection with phase changes.
The additional buoyancy flux due to phase changes and latent heat release 
can be quantified in terms of the parameter:
\begin{equation}
Q=\frac{1}{H}\int_0^H  Nu_B(z)\, \mbox{d}z - Nu_D\,.
\label{nud2}
\end{equation}
The upper and lower bounds on the buoyancy flux (\ref{inequality}) implies that the 
enhancement factor $Q$ is itself bound by $1 \le Q \le Ra_M/Ra_D$.   
Figure \ref{fig4} shows $Q$ as a function of $CSA$ and $Ra_D$.
The sensitivity of $Q$ to $CSA$ at a given value of $Ra_D$ is complex. 
On the one hand, high value of $CSA$ implies more water and a deeper saturated layer. On the other hand, 
in our experimental set-up  with $SSD = 0$ and constant $Ra_D$,  any increase of $CSA$ is connected
with a decrease of $Ra_M$. We  observe 
here a maximum 
of $Q$ at $CSA=0.44$. This is the case where in Fig. \ref{fig3} the largest amplitudes for $\langle u_zB(z)\rangle$ 
are observed (see dashed line in panel (c)).
\begin{figure}
\begin{center}
\includegraphics[width=8.5cm]{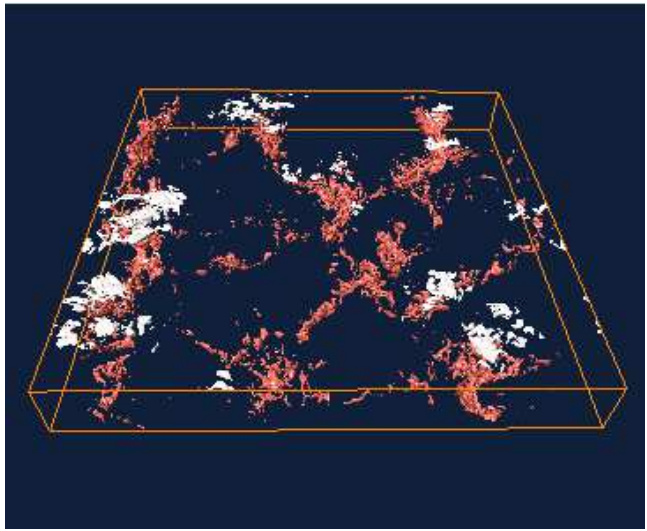}
\caption{Snapshot of the instantaneous cloud distribution (white isosurface) in combination
with the updrafts (red isosurface). The white isosurface is the cloud boundary with $q_l=0$.
The red isosurface is for $u_z\ge 0.23 U_f$ or $1.5\,u_{rms}$. Data are
for Run 14 with $CSA=0.26$. } 
\label{fig5}
\end{center}
\end{figure}

\subsection{Cloud cover}
The phase changes in the convective turbulence are associated with the appearance and disappearance of 
clouds. They are defined as those sites where the liquid water mixing ratio $q_l({\bf x},t)>0$. Translated into our
framework this corresponds with 
\begin{equation}
q_l({\bf x},t)=M({\bf x},t)-[D({\bf x},t)-N_s^2 z]>0\,.
\label{cl1}
\end{equation}
The cloud boundary is given by $q_l=M-D+N_s^2 z=0$. Depending on $CSA$ and both Rayleigh numbers 
this is a simply connected isosurface or a collection of disconnected isosurfaces. The latter case is illustrated
in Fig. \ref{fig5}. The white isosurfaces $q_l=0$ display isolated clouds. They are correlated with strong updrafts
as illustrated by the red isosurfaces for $u_z\ge 0.23 U_f$. Warm air rises up and expands adiabatically such 
that the temperature decreases and condensation sets in. 

\begin{figure}
\begin{center}
\includegraphics[width=11cm]{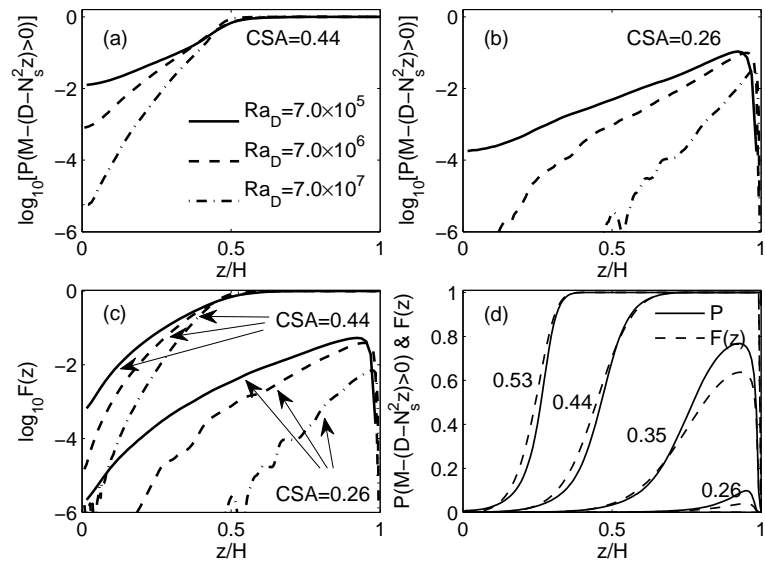}
\caption{Measures of the cloud cover. (a) Probability $P$ to find clouds in a layer at height $z$ as a function
of $Ra_D$. Data are for  $CSA=0.44$ . (b) Same as in (a) for $CSA=0.26$. The corresponding dry 
Rayleigh numbers for both figures are given in the legend of the left figure. (c) Function $F(z)$ as a function 
of $Ra_D$ and $CSA$. Line styles correspond with (a). (d) Comparison of  $P$ and $F(z)$ for runs 1 to  5. 
Data for run 5 at $CSA=0.17$ coincide with the axis. } 
\label{fig6}
\end{center}
\end{figure}

Figures \ref{fig6} (a) and (b) display the probability to find clouds at height $z$ in the slab as
a function of $Ra_D$ and $CSA$ in a semi-logarithmic plot. The formation of clouds is less probable when $Ra_D$
is increased. Reasons could be the stronger filamentation of the turbulent patches and the decreased velocity 
fluctuations which are in line with an
increase of the Reynolds number of the turbulent flow. For $CSA=0.53$ and 0.44 the cloud layer is closed 
for all $Ra_D$  which is in line with $P(z | q_l\ge 0)=1$. For $CSA=0.35$, 0.26 and 0.17, a broken cloud layer with isolated
clouds can be observed. While the former could correspond with a stratocumulus-like convection regime, the latter 
could correspond with a cumulus-like regime. Note also that layer remains basically dry for the smallest $CSA$.
           
The presence of clouds is also related to the enhancement of the buoyancy flux. We 
define a saturation fraction $F(z)$ 
for the buoyancy flux which follows from (\ref{inequality}) and is given by 
\begin{equation}
F(z) = \frac{\langle u_zB(z)\rangle - \langle u_zD(z)\rangle}{\langle u_zM(z)\rangle-\langle u_zD(z)\rangle}.
\label{fz}
\end{equation}
This saturation fraction is such that in a fully saturated environment $F(z) = 1$, while in an unsaturated environment, 
we have $F(z) = 0$.  Figures \ref{fig6} (c) and (d) replot the same data sets for $F(z)$ and in panel (d) a direct 
comparison with $P(z | q_l\ge 0)$ is provided. On the basis of the data, we can conclude that both measures collapse quite well for fully 
saturated or unsaturated layer, but the case $CSA = 0.35$ indicates there are some significant departures in partially 
saturated layers.  Finally, one can define now the vertical fraction of the layer that is partially saturated and unsaturated 
as $0.05<F(z)<0.95$ (see Fig. \ref{fig6} (d)). This fraction is biggest for the runs at $CSA=0.35$ as shown already in Fig. \ref{fig3} 
(b). In this case, the asymmetry between saturated ascents and unsaturated descents can be expected to be largest.  
Consequently, the largest asymmetries of the vertical velocity fluctuations can be built up. The peak at exactly 
the same $CSA$ value in both panels of Fig. \ref{fig3} supports our conclusion and closes the loop between
buoyancy transfer, cloud cover and vertical flow asymmetry.  

\section{Summary and Conclusions}
We have presented a shallow moist convection model with a linear equation of state for the 
thermodynamics of phase changes. This model which contains five dimensionless parameters 
is discussed in a three-dimensional subspace due to fixed Prandtl number and Surface Saturation 
Deficit ($SSD$). The most important simplification which reduces the complexity 
is the assumption of a local thermodynamic equilibrium. Several key physical processes, such as 
the formation of precipitation and  the existence of supersaturation are thus omitted. The model 
nevertheless captures the fundamental interactions between phase transition and dynamics. Phase 
changes cause an asymmetry of the vertical velocity fluctuations when the amount of water that 
can be condensed (parameter $CSA$) is sufficiently large. Similar to the dry convection case,
the correlations between vertical velocity and buoyancy are used to quantify the amount of additional 
buoyancy flux due to condensation and related latent heat release. Furthermore, this correlation can 
be connected with the cloud cover in the layer. 

The studies in this simplified setting provide thus a basis for possible parameterizations of cloud impact 
in large-scale models. In particular, determining the factors that control cloud fraction 
is a central issue in climate modeling, as small changes in cloud cover can dramatically affect the 
amount of energy received and emitted by the atmosphere. We found here that the Rayleigh number 
has a direct impact on the cloud cover, which should be a cause of concern, as the Rayleigh number in 
our numerical simulations ($Ra_D = 7\times 10^7$) is significantly smaller than its typical atmospheric 
value, $Ra \approx 10^{18}$--$10 ^{22}$. Nevertheless, the idealized moist Rayleigh-B\'{e}nard convection 
provides an important test for our understanding of clouds and of their sensitivity to environmental parameters.

\begin{acknowledgements}
We thank the DEISA Consortium (www.deisa.eu), co-funded through the EU
FP6 project RI-031513 and the FP7 project RI-222919, for support within
the DEISA Extreme Computing Initiative. JS is supported by 
DFG grants SCHU 1410/9-1 and SCHU 1410/5-1 and  OP by NSF grant ATM-0545047.
\end{acknowledgements}

\end{document}